\documentclass[aps,prl,floatfix,twocolumn,footinbib]{revtex4-1}
\usepackage{epsfig}
\usepackage{bm}
\usepackage{amssymb}
\usepackage{graphicx}
\usepackage{epstopdf}
\usepackage{amsmath}
\usepackage{natbib}
\usepackage{hyperref}
\usepackage{color}

\begin{document}
\title{Orbital degree of freedom  in artificial electron lattices on metal surface}
\author{Liang Ma}
\thanks{These two authors contributed equally}
\affiliation{School of Physics and Wuhan National High Magnetic Field Center,
Huazhong University of Science and Technology, Wuhan 430074,  P. R. China}
\author{Wen-Xuan Qiu}
\thanks{These two authors contributed equally}
\affiliation{School of Physics and Wuhan National High Magnetic Field Center,
Huazhong University of Science and Technology, Wuhan 430074,  P. R. China}
\author{Jing-Tao L\"{u} }
\email{jtlu@hust.edu.cn}
\affiliation{School of Physics and Wuhan National High Magnetic Field Center,
Huazhong University of Science and Technology, Wuhan 430074,  P. R. China}
\author{Jin-Hua Gao}
\email{jinhua@hust.edu.cn}
\affiliation{School of Physics and Wuhan National High Magnetic Field Center,
Huazhong University of Science and Technology, Wuhan 430074,  P. R. China}

\begin{abstract}
Orbital degree of freedom plays a fundamental role in condensed matter physics.
Recently, a new kind of artificial electron lattice has been realized in
experiments by confining the metal surface electrons with adsorbed molecular
lattice. A most recent example is the Lieb lattice realized by CO adsorption on
Cu(111) surface [M. R. Slot, \emph{et al}., Nat. Phys. \textbf{13}, 672
(2017)].  The Lieb lattice is of special interest due to its flat band physics.
Here, by first-principles calculations, muffin-tin potential model and tight
binding model, we demonstrate that, the high energy states observed in the
experiment actually correspond to the artificial $p$-orbitals of the electron
lattice. Our numerical results, together with the experimental observation,
show that artificial $p$-orbital fermionic lattice has already been realized in
solid state system. This opens a new avenue to investigate the orbital degree
of freedom in a controllable way.
\end{abstract}
\maketitle

The orbital degree of freedom, which refers to the orbital degeneracy and
orientational anisotropy, is a fundamental attribute of the Bloch electrons in
crystal, in addition to their charge and spin. When coupled to the
charge and spin, orbital degree of freedom of electrons can give rise to many
important phenomena, such as metal-insulator
transition, superconductivity, and colossal magneto-resistance\cite{nagaosa2000,hotta2006}.
However, understanding the orbital physics in real materials is still a big
challenge due to some realistic reasons, e.g, the constraints of materials, the
coupling among multiple degrees of freedom and the lack of controllability.

Utilizing artificial quantum systems is a promising way to study the
orbital physics. The most successful example is the cold atoms in optical
lattices, where atoms can be excited into higher orbital states of the optical
lattice. In last decade,  $p$-orbital related novel quantum states in optical
lattices have been intensively studied (see, for example,
Ref.~\onlinecite{li2016} and the references therein). For example,  for
$p$-band bosons, an unconventional BEC has been realized in
experiment\cite{pboson2007,bec2011,wubec2009}.  However, an experimental
realization of $p$-band fermions has not been reported in any artificial
quantum lattices so far.

Recently, a new kind of artificial two dimensional electron lattice has been
realized in experiments. The first example is the molecular
graphene\cite{nature2012,lin2014}, where the Cu surface electrons, a perfect
two dimensional electron gas (2DEG) with $k^2$ dispersion, is confined in a
honeycomb lattice by lateral periodic potential induced by adsorbed
molecules. Even more exciting is the realization of artificial Lieb lattice\cite{qiu2016, slot2017,drost2017},
that has not been found in
natural materials and is of special interest due to its flat band physics.
These pioneering works demonstrate that artificial electron lattice on metal
surface could be a promising solid-state quantum simulation platform.

An interesting issue in the recent experiment\cite{slot2017} is that some
high-energy electron states with complex local density of states (LDOS) pattern
are observed in the scanning tunnelling microscope (STM), which are drastically
different from the low energy states. Here, combining first-principles simulation based on
Density Functional Theory (DFT), muffin-tin potential model and tight-binding calculations,
we demonstrate that these high energy states originate from artificial
$p$-orbitals of the electron lattice. Through carefully comparison of our theoretical
calculation with the experimental observation, we show that the experiment in
Ref.~\onlinecite{slot2017} has actually realized an artificial
$p$-orbital square (and Lieb) electron lattice in
solid state system. It may give the first example of the two dimensional
$p$-orbital fermion lattice in artificial quantum systems, and offers
an ideal solid state  platform to study orbital physics in a controllable way.

\begin{figure*}
\centering
\includegraphics[width=18cm]{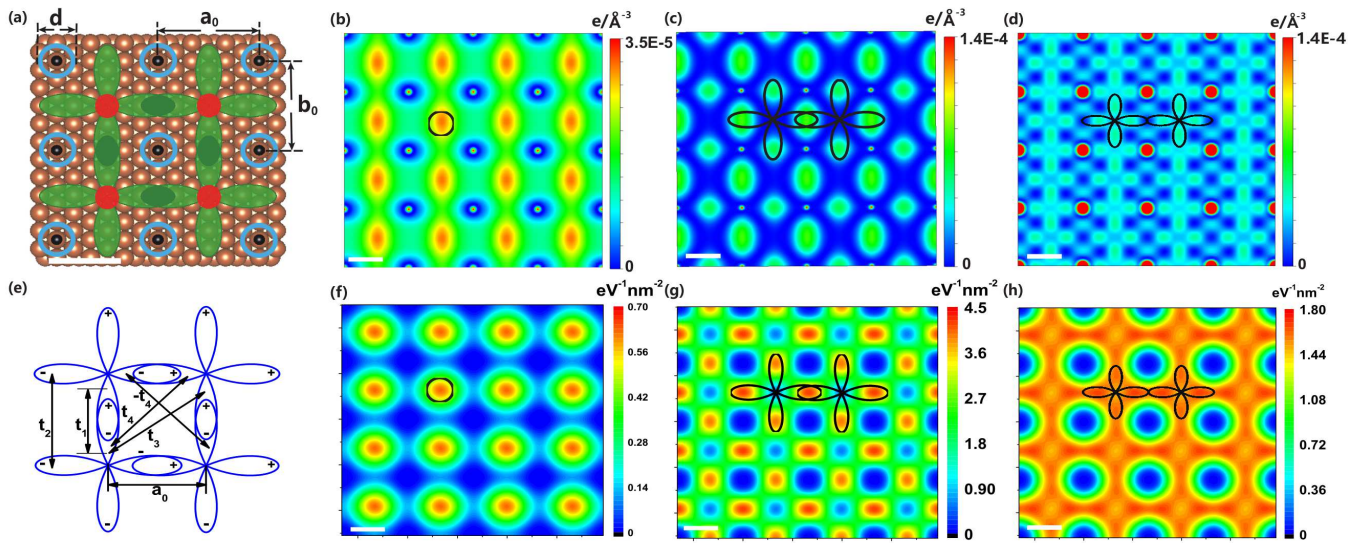}
\caption{(Color online) (a) Schematic  of the artificial electron square lattice on Cu(111) surface. The red discs are the $s$-orbitals and the green lobes denote the $p$-orbitals. The lattice vectors are $a_0 \approx 1.33$ nm and $b_0 \approx 1.28$ nm, respectively.  (b),(c),(d) are the DFT simulated STM images, where the energy region, with respect to the Fermi level, are $-0.3\sim-0.2$ eV, $-0.1\sim-0.08$ eV and $0.55\sim0.65$ eV, respectively.  (e) Schematic of the $p$-orbital tight binding model on square lattice. (f),(g),(h) are the LDOS calculated from the muffin-tin potential model, where the energy  are $1.14$ eV, $1.47$ eV and $2.15$ eV, respectively. $a_0=1.33$ nm, $b_0=1.28$ nm, $U_0=9$ eV, $d=0.5$ nm. Scale bars, 1 nm.
}
\label{fig1}
\end{figure*}

Let us start with the square electron lattice on metal surface. The structure
is shown in Fig. \ref{fig1} (a), where the CO molecules
(black balls) are arranged on the Cu(111) surface to form a square lattice.
Note that, due to the geometry of the Cu(111) surface, the lattice constant (CO
molecule lattice) in $x$ direction $a_0$ is slightly different from that in $y$
direction $b_0$. For simplicity, we ignore this difference in the muffin-tin potential model
unless specified otherwise. Here, each CO molecule exerts a repulsive potential on the Cu
surface electrons, which can be described by a muffin-tin potential $U(r)$ [see
Fig. \ref{fig1} (a), $U_0>0$ inside the blue  circles and zero elsewhere, $d$
is the diameter of circular potential]. The Hamiltonian of the Cu surface
states now is
\begin{equation}
\label{hamitonian}
H_{\rm Cu}=\frac{\hbar^2}{2m^*}\bm{k}^2 +U(r)
\end{equation}
where $m^*=0.38m_e$ is the effective mass of the Cu surface electrons, and
$U_0=9$ eV, $d=0.5$ nm are reasonable values for the muffin-tin potential of
CO/Cu(111) system\cite{lishuai2016,dft2014}. It is actually an anti-dot
lattice, since the surface electrons under the CO molecules are depleted by the
repulsive potential $U(r)$.  Consequently, surface electrons are forced into a
square lattice, where the electron sites are in the center of the
squares formed by four adjacent CO molecules. Theoretically, the corresponding
band structure can be obtained by solving the Hamiltonian of the muffin-tin
potential model with plane wave basis. Here, the LDOS of surface
electrons is the most important quantity, which actually represents the
electron wave functions and can be directly measured by STM. Theoretically,
the LDOS can  be calculated from
\begin{equation}
\label{LDOS}
\textrm{LDOS}(r,\varepsilon)=\sum_{nk\sigma}|\phi_{nk\sigma}(r)|^2\delta(\varepsilon-\epsilon_{nk\sigma}),
\end{equation}
where $\phi_{nk\sigma}(r)$ is the wave function of surface electron, $n$, $k$
and $\sigma$ are the indices of band, momentum,  and spin, respectively.

In addition to the muffin-tin potential model, we perform DFT calculations
to direct simulate the experimentally realized system\cite{dft2014}.
We use the Vienna $Ab$-$initio$ Simulation Package (VASP), which is based on the projector-augmented wave method and a
plane wave basis set\cite{vasp}.  We choose the Perdew-Burke-Ernzerhof (PBE)
version of the generalized gradient approximation\cite{gga}, and the energy
cutoff is 400 eV. A four-layer slab is used to model the Cu(111) surface and a
10 \AA \ vacuum region separates slabs between nearest supercells to avoid their
interaction. The positions of the CO molecules and the top layer Cu
atoms are optimized in a smaller supercell and used in present calculation.
We can directly simulate the STM
image (LDOS of surface electron states) using the Tersoff-Hamann scheme\cite{tersoff1983} from
the DFT electronic structure.

\begin{figure}
\centering
\includegraphics[width=8.5cm]{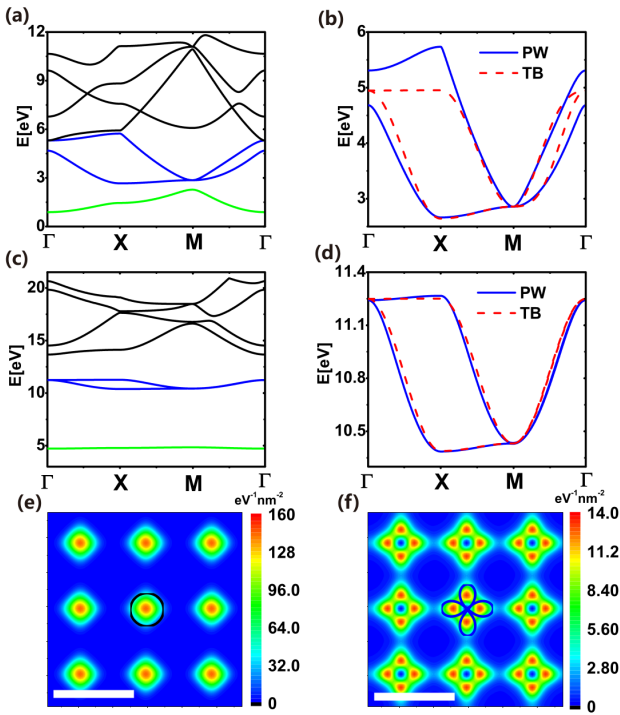}
\caption{(Color online) (a) Energy bands of the electron square lattice, where green lines are the $s$-bands, blue lines are the $p$-bands. $U_0=9$ eV, $d=0.5$ nm, $a_0=0.95$ nm. (b) Fitting the $p$-bands in (a) (blue, solid) with a $p$-orbital tight binding model on square lattice (red, dashed). PW means the muffin-tin model with plane wave basis. [$t_1,t_2,t_3,t_4$]=[0.55, -0.0275, 0.0131, 0.0963] eV. $\varepsilon_{p_x}=\varepsilon_{p_y}=3.85$ eV. (c) Energy bands of the electron square lattice in the atomic limit. $a_0=0.95$ nm,  $U_0=15$ eV and $d=0.9$ nm. (d) Fitting the $p$-bands in (c) (blue, solid) with the tight binding model (red, dashed). [$t_1,t_2,t_3,t_4$]=[0.21, -0.0055, 0.0026, 0.0014] eV, $\varepsilon_{p_x}=\varepsilon_{p_y}=10.83$ eV. (e) LDOS of the $s$-bands in atomic limit, $E=4.8$ eV. (f) LDOS of the $p$-bands in atomic limit at $E=10.8$ eV. Scale bars, 1 nm.}
\label{fig2}
\end{figure}



First, we show that the LDOS of this
artificial square lattice observed in experiment can be well reproduced by the
both muffin-tin potential model and the first-principles calculation. The numerical results
are summarized in Fig.\ref{fig1}, where Fig. \ref{fig1} (b),(c),(d) are results of
first-principles calculation, and Fig. \ref{fig1} (f),(g),(h) are those from
the muffin-tin potential model. Basically, three typical surface
electron LDOS patterns at different energies have been observed in experiment:
(1) at low energy [Fig. \ref{fig1} (b),(f)],  the electron states are localized
around the lattice sites, so we get a square lattice pattern; (2)
for a high energy state [Fig. \ref{fig1} (c),(g)], the electron states are now
located in between the lattice sites; (3) increasing the energy further, though the
electron states are still mainly in between the lattice sites, some fine
structures of the LDOS pattern appear, i.e., we find a LDOS node
between two adjacent lattice sites [Fig. \ref{fig1} (d),(h)].  Both DFT and
the muffin-tin potential model nicely reproduce the experimental observations
[Fig.~$4$ (e), (f), (g) in Ref.~\onlinecite{slot2017}].


The main focus of this work is to understand the physical origin of these complex
electron LDOS (wave functions) pattern. Our  viewpoint here is that, the
low energy states are from the $s$-band of the electron square lattice,
while the high energy states are  from the $p$-bands.  In other words, even for
this anti-dot lattice, the orbital degree of freedom is still a valid concept.
Here, each electron lattice site could be considered as an artificial atom in
two dimensions with various atomic orbitals, such as $s$-and $p$-orbitals (only
$p_x$ and $p_y$ here since it is a two dimensional system).  These artificial
orbitals form the corresponding energy bands due to hopping between
them. Thus, Fig.~\ref{fig1} (b-d) and (f-h) actually show the real space distribution of these
orbital bands.
Importantly, in Fig. \ref{fig1} (c),  we see that the $p$-orbitals
of two neighboring sites form a $\sigma$ bond, which is localized in between
the sites. Thus the STM measurements actually have
observed the artificial $p$-orbital $\sigma$ bonds. Meanwhile, at higher
energy, the $p$-orbitals form  anti-bonding $\sigma^*$ bonds, which have
function nodes in between the artificial atoms. This is the origin of the LDOS
nodes observed in the STM measurement.

Given the success of our muffin-tin potential model,
we further use it to calculate the energy bands of the
square lattice in order to prove the above physical picture.
We show that the calculated
band structure can be well understood from the orbital point of view. The
calculated bands are shown in Fig. \ref{fig2} (a), where we set $a_0=b_0=0.95$ nm and $U_0=9$ eV, $d=0.5$ nm are the parameters for CO/Cu(111) system.
Note that $a_0$ is the distance between molecules which is tunable in
experiment. Different $a_0$ gives different hopping amplitude, but the
corresponding band shape is similar. We see that there is one band with the
lowest energy (green solid line) gapped from the others. This is the
$s$-band, because in  square lattice each site contributes one $s$-orbital and
these $s$-orbitals form one $s$-band with lowest energy. The $s$-orbitals are
around the lattice sites, and thus should give an electron square lattice.  In
our calculation, in the energy interval of the $s$-bands, the corresponding
LDOS are all like Fig. \ref{fig1} (b), which show a clear square lattice
LDOS pattern and are consistent with the experimental observation.  Note that
the $s$-band from the muffin-tin potential calculation can be well described by a single
band tight-binding model on square lattice. Similarly, the
upper two bands above the $s$-band (blue solid lines) should be the $p$-bands,
because each 2D atom has two $p$-orbitals, i.e. $p_x$ and $p_y$. We use a
two-band ($p_x$, $p_y$) tight binding model on square lattice, which is illustrated in Fig. \ref{fig1} (e), to
fit the $p$-bands. The tight binding
Hamiltonian is
\begin{equation}
H(k)=\left(
\begin{array}{cc}
H_{p_x}&V_{{p_x}{p_y}}\\
V_{{p_y}{p_x}}&H_{p_y}
\end{array}
\right),
\end{equation}
where
\begin{equation}
\begin{split}
H_{p_x}=&\epsilon_{p_x}+2t_1 \cos(k_x a_0)+2t_2 \cos(k_y a_0)\\
&+4t_3 \cos(k_x a_0)\cos(k_y a_0),   \\
H_{p_y}=&\epsilon_{p_y}+2t_1 \cos(k_y a_0)+2t_2 \cos(k_x a_0)\\
&+4t_3 \cos(k_x a_0)\cos(k_y a_0),\\
V_{{p_x}{p_y}}=&-4t_4 \sin(k_x a_0)\sin(k_y a_0).
\end{split}
\end{equation}
Here, $V_{{p_x}{p_y}}=V_{{p_y}{p_x}}$.  The nearest neighbour hopping $t_1$,
$t_2$ and the next nearest neighbour hopping $t_3$, $t_4$ are considered, which are
illustrated in Fig. \ref{fig1} (e). The fitting is quite well as shown in
Fig.\ref{fig2} (b), except that the tight binding model can not reproduce the
gap at the $\Gamma$ point. We attribute this difference to the band overlap
with the $d$-bands.  As can be seen in Fig. \ref{fig2} (a), there is no gap
between the $p$- and $d$-bands. This means that there should be some overlap
between them, which is not included in the tight binding model.

\begin{figure*}
\centering
\includegraphics[width=18cm]{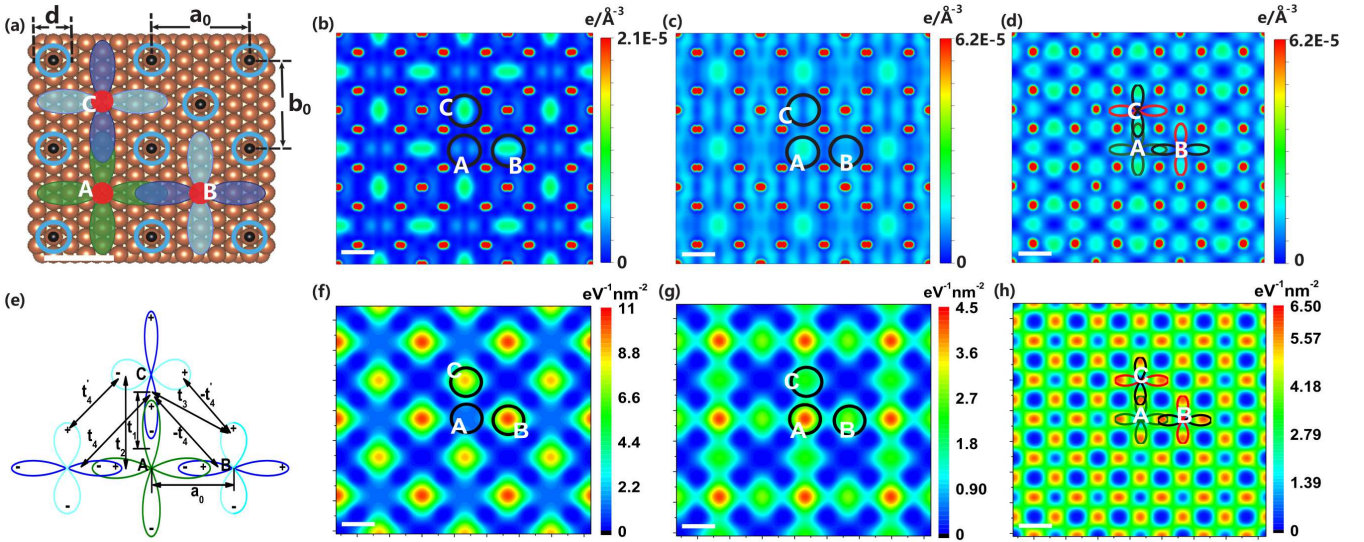}
\caption{(Color online) (a) Schematic of Lieb lattice. (b),(c),(d) are the DFT simulated STM image (LDOS), where the energy regions are $-0.5\sim-0.4$ eV, $-0.2\sim-0.1$ eV and $0.2\sim0.3$ eV, respectively. $a_0=1.33$ nm, $b_0=1.28$ nm. (e) Schematic of the $p$-orbital tight binding model on Lieb lattice. The $p$-orbitals with different energy are denoted with different colors.  (f),(g),(h) are the calculated LDOS with muffin-tin potential model, the energy of which are $0.61$ eV, $0.75$ eV, and $1.29$ eV, respectively. $a_0=1.33$ nm, $b_0=1.28$ nm, $U_0=9$ eV, $d=0.5$ nm. Scale bars, 1 nm.
}
\label{fig3}
\end{figure*}

In order to make the $p$-orbital picture more clear, we further consider an
extreme case  in the muffin-tin potential model, where we set $a_0=0.95$ nm, and use a fictional muffin-tin potential
$U_0=15$ eV and $d=0.9$ nm.   With given $a_0$ and very large values of $U_0$
and $d$ ($a_0>d$),  the hopping between adjacent sites is greatly suppressed,
so that the artificial atoms are nearly isolated. Thus, it is just the atomic limit.  The band structure in this
atomic limit is given in Fig. \ref{fig2} (c). It is now obvious that the two
$p$-bands are separated from the higher $d$-bands, i.e. there is no band
overlap any more.  As a consequence, the gap of the $p$-bands at the $\Gamma$
point disappears . The fitting of $p$-bands  is given in Fig. \ref{fig2} (d).
The tight binding model now can give a perfect description of the $p$-bands.

We plot the LDOS pattern in the atomic limit in Fig.
\ref{fig2} (e) and (f). At low energy [Fig. \ref{fig2} (e)], we see that the
isotropic $s$-orbitals  are well separated from each other.  Increasing the
energy to the $p$-band region, we get an LDOS pattern in
Fig. \ref{fig2} (f), resembling the shape of isolated $p$-orbitals. All the discussions above give a clear
illustration of the $p$-orbital picture in the square lattice.

Another important issue is the position of Fermi level relative to the
$p$-bands, which depends on the square lattice constant $a_0$\cite{nature2012,lishuai2016,qiu2016}.
We can give an estimation about
the value of $a_0$ to access the $p$-bands in square lattice.  In square
lattice, the number of electrons in each site is about $N_e a^2_0$, where $N_e$
is the electron density of the metal surface. For Cu(111) surface, $N_e$ is
about $0.72$ nm$^{-2}$. Here, we assume that the adsorbed CO molecules do not
modify the surface electron number of Cu. It is reasonable for the experimental
situation, since we do not expect that one hundred CO molecules can change the
Fermi level of the bulk Cu crystal. To access the $p$-bands, $a_0$ should be in the
region $1.7$ nm $< a_0 < 2.9$ nm. This estimation is consistent with the
experimental observations. In experiment\cite{slot2017}, two square lattices
are realized. The $a_0$ of large one is about $2.56$ nm ($b_0 \approx 2.66$
nm), two times larger than that of the small one. So, according to the
estimation above, the Fermi level of the large square lattice should be at the
$p$-bands, while that of the small lattice is at the $s$-bands. In the STM
measurement, for the large square lattice,  the LDOS patterns of the $s$-band
and bonding $p$-band are found below the Fermi level, while that of the
anti-bonding $p$-band is above the Fermi level. This is consistent with our estimation.
In contrast, for the
small square lattice,  no matter with positive or negative bias voltage, only the LDOS pattern of the $s$-band is observed ,
indicating that the Fermi level is at the $s$-band. In order to access the
$p$-band,  more larger bias voltage is needed for the small square lattice case.


\begin{figure*}[t]
\centering
\includegraphics[width=18cm]{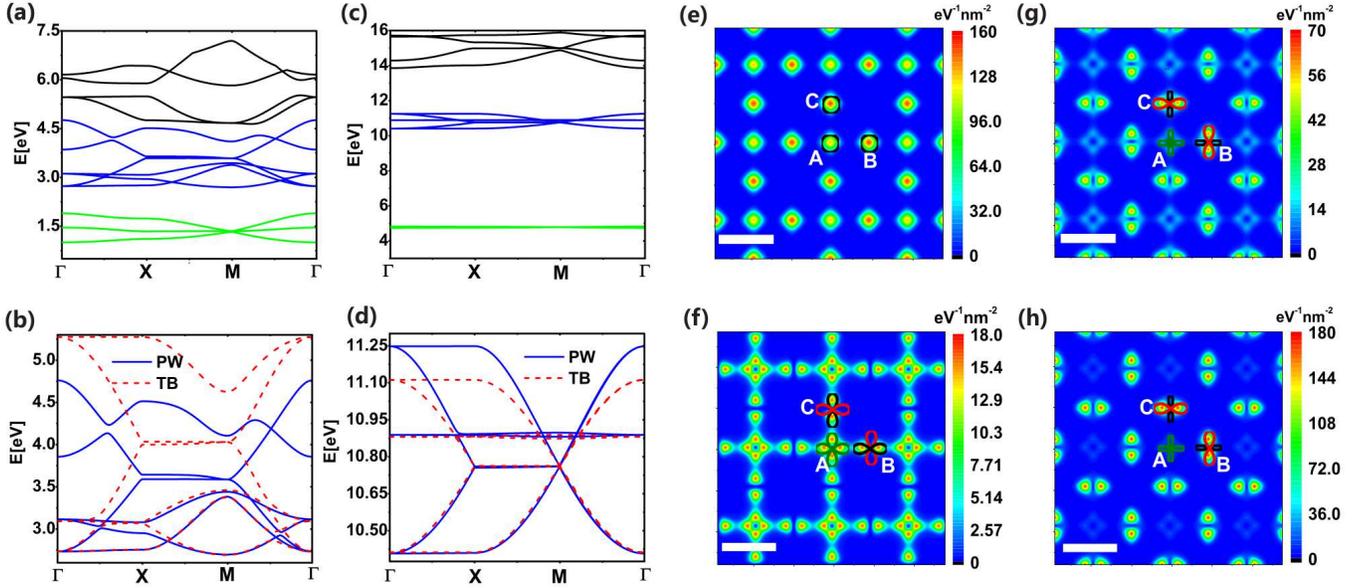}
\caption{(Color online) (a) Energy bands of the Lieb lattice. $a_0=0.95$ nm, $d=0.5$ nm, $U_0=9$ eV. (b) Fitting the $p$-bands in (a) (blue, solid) with the $p$-orbital tight binding model (red, dashed) on Lieb lattice. [$t_1$,$t_2$,$t_3$,$t_4$,$t'_4$]=[0.63,-0.0345,0.011,0.1563,0.096] eV.  $\varepsilon_{\textrm{A}p_x} = \varepsilon_{\textrm{A}p_y} = 4.03$ eV, $\varepsilon_{\textrm{B}p_x} =\varepsilon_{\textrm{C}p_y} = 4$ eV, $\varepsilon_{\textrm{B}p_y} =\varepsilon_{\textrm{C}p_x} = 3.075$ eV. (c) Energy bands of the Lieb lattice in the atomic limit. $a_0=0.95$ nm, $d=0.9$ nm, $U_0=15$ eV. (d) Fitting the $p$-bands in (c) (blue, solid) with the $p$-orbital tight binding model (red, dashed) on Lieb lattice.  [$t_1$,$t_2$,$t_3$,$t_4$,$t'_4$]=[0.175,-0.0055,0.0026,0.0016,0.0019] eV.  $\varepsilon_{\textrm{A}p_x} = \varepsilon_{\textrm{A}p_y} = 10.761$ eV, $\varepsilon_{\textrm{B}p_x} =\varepsilon_{\textrm{C}p_y} = 10.763$ eV, $\varepsilon_{\textrm{B}p_y} =\varepsilon_{\textrm{C}p_x} = 10.881$ eV. (e)-(h) The LDOS of the Lieb lattice in the atomic limit. The energies are taken at $4.8$ eV, $10.6$ eV, $10.8$ eV, $10.9$ eV, respectively. Scale bars, 1 nm.
}
\label{fig4}
\end{figure*}


Now, we turn to the Lieb lattice.  As shown in Fig. \ref{fig3} (a), there are three lattice
sites (A, B,C) in each unit cell of a Lieb lattice. Thus, there are three $s$-bands
with lowest energy,  of which the middle one is a flat band if only the nearest
neighbor hopping is considered. A unique property of this flat band
is its special LDOS pattern. The electrons in the flat band  are only
localized at the B and C sites, while the electrons in other two $s$-bands mainly
distribute around the A sites. This unique LDOS pattern actually reflects the
flat band localization phenomenon in Lieb lattice, and has been observed in
different systems\cite{qiu2016,slot2017,liebprl2015a,liebprl2015b}.

We first show that, for the $s$-bands, the special LDOS pattern of the flat
band (the flat band localization phenomenon) can be well reproduced by the
our calculation. The DFT results are shown in Fig. \ref{fig3} (b) and (c), while the corresponding
muffin-tin results are given in Fig. \ref{fig3} (f) and (g) as a comparison.
In Fig. \ref{fig3} (b), it is seen that the
electron states are absent around the A site, which is just the special LDOS
pattern of the Lieb flat band.  Increasing the energy to the upper $s$-band,
the electrons should distribute mainly  around the A site, instead of the B and
C sites. This is observed in Fig. \ref{fig3} (c). The muffin-tin potential
model can also reproduce these features, as shown in Fig.\ref{fig3} (f),(g).

The $p$-orbital picture also works well for the high energy states in the Lieb
lattice.  It is not surprising because a square lattice can be changed into a
Lieb lattice by removing one of the four sites.  The illustration of the
$p$-orbitals in Lieb lattice is given in Fig. \ref{fig3} (e). Similar to the
square lattice, we expect that the $p$-band electrons will mainly distribute
in between the sites, due to the formed $p$-orbital $\sigma$ bond, as was found
in the experiment. The calculated LDOS confirms our expectation. We can see that, the experimental
observation, the first-principles calculation [Fig. \ref{fig3} (d)], and the
muffin-tin model [Fig. \ref{fig3} (h)] all coincide well with the $p$-orbital
picture.  The main distinction from the square lattice is that the $p$-orbitals
at the A, B, C sites are now inequivalent, due to their different geometric
environment. They thus have different on-site energies, which are denoted with
different colors in Fig.  \ref{fig3} (e).

We plot the band structure of the Lieb lattice calculated
from the muffin-tin potential model with plane wave basis in Fig. \ref{fig4}.
We first consider a normal situation [Fig.
\ref{fig4} (a)], where $a_0= 0.95$ nm, $U_0=9$ eV.  In this case, we see that
the lowest three bands (green solid lines) are $s$-bands of
Lieb lattice. Considering the $p$-orbitals,  there should be six $p$-bands  above
the three $s$-bands. This is shown in  Fig. \ref{fig4} (a) as blue solid lines,
separated from both the lower $s$-bands and
the upper $d$-bands. These $p$-bands can  be qualitatively interpreted  by a
$p$-orbital tight binding model on Lieb lattice [see in Fig. \ref{fig3} (e)]. In
Fig. \ref{fig4} (b), we show the tight binding fitting of the band structure.
At low energy, the tight binding model works quite well, but it
can not describe the top most two $p$-bands. Similar to the square lattice, we
attribute this discrepancy to the influence of the upper $d$-bands. The
corresponding LDOS of the $p$-bands are like Fig. \ref{fig3} (d),(h), where the
$p$-orbital $\sigma$ bonds  are shown clearly.

We now consider the atomic limit of the Lieb lattice ($U_0=15$ eV, $d=0.9$ nm,
$a_0=0.95$ nm) to support our $p$-orbital picture.
In the atomic limit, the hopping between adjacent sites is greatly
suppressed. The bands of the atomic limit are shown in in Fig. \ref{fig4} (c).
In this case, the $p$-bands are far from the $s$- and $d$-bands.
In Fig. \ref{fig4} (d), we also use the $p$-orbital tight binding model to fit these $p$-bands.
Now the agreement is much better.  Since the next nearest neighbor  hopping becomes very tiny here,  some $p$-orbitals on
the B and C sites  now form  dangling bonds, which results in degenerate flat
$p$-bands. The LDOS in atomic limit are also given in Fig. \ref{fig4}. At low
energy, the $s$-orbitals are around the lattice sites, and form a Lieb lattice
[Fig. \ref{fig4} (e)]. Note that, because the hopping is tiny in
the atomic limit, the artificial atoms are nearly isolated. Thus, the unique
LDOS pattern of flat band we mentioned above can not be observed here.
Continuously increasing the energy, the LDOS can sequentially show different
$p$-orbitals  as illustrated in Fig. \ref{fig4} (f), (g) and (h). As we mentioned
above, this is because the energy of the  $p$-orbitals in Lieb lattice are
different.

In summary, we theoretically demonstrate that the high energy states in the
artificial electron square (Lieb) lattice,  as observed in the recent STM
experiment, are from the $p$-bands of the artificial atom confined in the
lattice. The orbital degree of freedom is still a valid concept in this
artificial electron lattice system. Our results suggest that, the electron
lattice realized in Ref.~\onlinecite{slot2017} may be the first artificial
$p$-orbital fermionic system in the solid state.  Compared with other
artificial quantum  systems, this kind of electron lattice on metal surface is
easy to manipulate, and   the electron states can be directly detected. Thus,
we believe that it is an ideal solid state platform to study orbital physics.
Finally, we comment that, the same physics applies to artificial antidot lattice
on 2DEG in semiconductor
heterostructures\cite{nanoletter2009,polini2013,west2009}.


\begin{acknowledgments}
J.H.G. and J.T.L. are supported by the National Natural Science Foundation of China (Grants No. 11534001, 61371015,11274129).
\end{acknowledgments}

\bibliography{pbandbib}

\begin{thebibliography}{21}%
\makeatletter
\providecommand \@ifxundefined [1]{%
 \@ifx{#1\undefined}
}%
\providecommand \@ifnum [1]{%
 \ifnum #1\expandafter \@firstoftwo
 \else \expandafter \@secondoftwo
 \fi
}%
\providecommand \@ifx [1]{%
 \ifx #1\expandafter \@firstoftwo
 \else \expandafter \@secondoftwo
 \fi
}%
\providecommand \natexlab [1]{#1}%
\providecommand \enquote  [1]{``#1''}%
\providecommand \bibnamefont  [1]{#1}%
\providecommand \bibfnamefont [1]{#1}%
\providecommand \citenamefont [1]{#1}%
\providecommand \href@noop [0]{\@secondoftwo}%
\providecommand \href [0]{\begingroup \@sanitize@url \@href}%
\providecommand \@href[1]{\@@startlink{#1}\@@href}%
\providecommand \@@href[1]{\endgroup#1\@@endlink}%
\providecommand \@sanitize@url [0]{\catcode `\\12\catcode `\$12\catcode
  `\&12\catcode `\#12\catcode `\^12\catcode `\_12\catcode `\%12\relax}%
\providecommand \@@startlink[1]{}%
\providecommand \@@endlink[0]{}%
\providecommand \url  [0]{\begingroup\@sanitize@url \@url }%
\providecommand \@url [1]{\endgroup\@href {#1}{\urlprefix }}%
\providecommand \urlprefix  [0]{URL }%
\providecommand \Eprint [0]{\href }%
\providecommand \doibase [0]{http://dx.doi.org/}%
\providecommand \selectlanguage [0]{\@gobble}%
\providecommand \bibinfo  [0]{\@secondoftwo}%
\providecommand \bibfield  [0]{\@secondoftwo}%
\providecommand \translation [1]{[#1]}%
\providecommand \BibitemOpen [0]{}%
\providecommand \bibitemStop [0]{}%
\providecommand \bibitemNoStop [0]{.\EOS\space}%
\providecommand \EOS [0]{\spacefactor3000\relax}%
\providecommand \BibitemShut  [1]{\csname bibitem#1\endcsname}%
\let\auto@bib@innerbib\@empty
\bibitem [{\citenamefont {Tokura}\ and\ \citenamefont
  {Nagaosa}(2000)}]{nagaosa2000}%
  \BibitemOpen
  \bibfield  {author} {\bibinfo {author} {\bibfnamefont {Y.}~\bibnamefont
  {Tokura}}\ and\ \bibinfo {author} {\bibfnamefont {N.}~\bibnamefont
  {Nagaosa}},\ }\href {\doibase 10.1126/science.288.5465.462} {\bibfield
  {journal} {\bibinfo  {journal} {Science}\ }\textbf {\bibinfo {volume}
  {288}},\ \bibinfo {pages} {462} (\bibinfo {year} {2000})}\BibitemShut
  {NoStop}%
\bibitem [{\citenamefont {Hotta}(2006)}]{hotta2006}%
  \BibitemOpen
  \bibfield  {author} {\bibinfo {author} {\bibfnamefont {T.}~\bibnamefont
  {Hotta}},\ }\href {\doibase 10.1088/0034-4885/69/7/R02} {\bibfield  {journal}
  {\bibinfo  {journal} {Rep. Prog. Phys.}\ }\textbf {\bibinfo {volume} {69}},\
  \bibinfo {pages} {2061} (\bibinfo {year} {2006})}\BibitemShut {NoStop}%
\bibitem [{\citenamefont {Li}\ and\ \citenamefont {Liu}(2016)}]{li2016}%
  \BibitemOpen
  \bibfield  {author} {\bibinfo {author} {\bibfnamefont {X.}~\bibnamefont
  {Li}}\ and\ \bibinfo {author} {\bibfnamefont {W.~V.}\ \bibnamefont {Liu}},\
  }\href {\doibase 10.1088/0034-4885/79/11/116401} {\bibfield  {journal}
  {\bibinfo  {journal} {Rep. Prog. Phys.}\ }\textbf {\bibinfo {volume} {79}},\
  \bibinfo {pages} {116401} (\bibinfo {year} {2016})}\BibitemShut {NoStop}%
\bibitem [{\citenamefont {M\"uller}\ \emph {et~al.}(2007)\citenamefont
  {M\"uller}, \citenamefont {F\"olling}, \citenamefont {Widera},\ and\
  \citenamefont {Bloch}}]{pboson2007}%
  \BibitemOpen
  \bibfield  {author} {\bibinfo {author} {\bibfnamefont {T.}~\bibnamefont
  {M\"uller}}, \bibinfo {author} {\bibfnamefont {S.}~\bibnamefont {F\"olling}},
  \bibinfo {author} {\bibfnamefont {A.}~\bibnamefont {Widera}}, \ and\ \bibinfo
  {author} {\bibfnamefont {I.}~\bibnamefont {Bloch}},\ }\href {\doibase
  10.1103/PhysRevLett.99.200405} {\bibfield  {journal} {\bibinfo  {journal}
  {Phys. Rev. Lett.}\ }\textbf {\bibinfo {volume} {99}},\ \bibinfo {pages}
  {200405} (\bibinfo {year} {2007})}\BibitemShut {NoStop}%
\bibitem [{\citenamefont {Wirth}\ \emph {et~al.}(2011)\citenamefont {Wirth},
  \citenamefont {\"{O}lschl\"{a}ger},\ and\ \citenamefont
  {Hemmerich}}]{bec2011}%
  \BibitemOpen
  \bibfield  {author} {\bibinfo {author} {\bibfnamefont {G.}~\bibnamefont
  {Wirth}}, \bibinfo {author} {\bibfnamefont {M.}~\bibnamefont
  {\"{O}lschl\"{a}ger}}, \ and\ \bibinfo {author} {\bibfnamefont
  {A.}~\bibnamefont {Hemmerich}},\ }\href {\doibase 10.1038/nphys1857}
  {\bibfield  {journal} {\bibinfo  {journal} {Nat. Phys.}\ }\textbf {\bibinfo
  {volume} {7}},\ \bibinfo {pages} {147} (\bibinfo {year} {2011})}\BibitemShut
  {NoStop}%
\bibitem [{\citenamefont {Wu}(2009)}]{wubec2009}%
  \BibitemOpen
  \bibfield  {author} {\bibinfo {author} {\bibfnamefont {C.}~\bibnamefont
  {Wu}},\ }\href {\doibase 10.1142/S0217984909017777} {\bibfield  {journal}
  {\bibinfo  {journal} {Mod. Phys. Lett. B}\ }\textbf {\bibinfo {volume}
  {23}},\ \bibinfo {pages} {1} (\bibinfo {year} {2009})}\BibitemShut {NoStop}%
\bibitem [{\citenamefont {Gomes}\ \emph {et~al.}(2012)\citenamefont {Gomes},
  \citenamefont {Mar}, \citenamefont {Ko}, \citenamefont {Guinea},\ and\
  \citenamefont {Manoharan}}]{nature2012}%
  \BibitemOpen
  \bibfield  {author} {\bibinfo {author} {\bibfnamefont {K.~K.}\ \bibnamefont
  {Gomes}}, \bibinfo {author} {\bibfnamefont {W.}~\bibnamefont {Mar}}, \bibinfo
  {author} {\bibfnamefont {W.}~\bibnamefont {Ko}}, \bibinfo {author}
  {\bibfnamefont {F.}~\bibnamefont {Guinea}}, \ and\ \bibinfo {author}
  {\bibfnamefont {H.~C.}\ \bibnamefont {Manoharan}},\ }\href {\doibase
  10.1038/nature10941} {\bibfield  {journal} {\bibinfo  {journal} {Nature}\
  }\textbf {\bibinfo {volume} {483}},\ \bibinfo {pages} {306} (\bibinfo {year}
  {2012})}\BibitemShut {NoStop}%
\bibitem [{\citenamefont {Wang}\ \emph {et~al.}(2014)\citenamefont {Wang},
  \citenamefont {Tan}, \citenamefont {Wang}, \citenamefont {Louie},\ and\
  \citenamefont {Lin}}]{lin2014}%
  \BibitemOpen
  \bibfield  {author} {\bibinfo {author} {\bibfnamefont {S.}~\bibnamefont
  {Wang}}, \bibinfo {author} {\bibfnamefont {L.~Z.}\ \bibnamefont {Tan}},
  \bibinfo {author} {\bibfnamefont {W.}~\bibnamefont {Wang}}, \bibinfo {author}
  {\bibfnamefont {S.~G.}\ \bibnamefont {Louie}}, \ and\ \bibinfo {author}
  {\bibfnamefont {N.}~\bibnamefont {Lin}},\ }\href {\doibase
  10.1103/PhysRevLett.113.196803} {\bibfield  {journal} {\bibinfo  {journal}
  {Phys. Rev. Lett.}\ }\textbf {\bibinfo {volume} {113}},\ \bibinfo {pages}
  {196803} (\bibinfo {year} {2014})}\BibitemShut {NoStop}%
\bibitem [{\citenamefont {Qiu}\ \emph {et~al.}(2016)\citenamefont {Qiu},
  \citenamefont {Li}, \citenamefont {Gao}, \citenamefont {Zhou},\ and\
  \citenamefont {Zhang}}]{qiu2016}%
  \BibitemOpen
  \bibfield  {author} {\bibinfo {author} {\bibfnamefont {W.-X.}\ \bibnamefont
  {Qiu}}, \bibinfo {author} {\bibfnamefont {S.}~\bibnamefont {Li}}, \bibinfo
  {author} {\bibfnamefont {J.-H.}\ \bibnamefont {Gao}}, \bibinfo {author}
  {\bibfnamefont {Y.}~\bibnamefont {Zhou}}, \ and\ \bibinfo {author}
  {\bibfnamefont {F.-C.}\ \bibnamefont {Zhang}},\ }\href
  {http://link.aps.org/doi/10.1103/PhysRevB.94.241409} {\bibfield  {journal}
  {\bibinfo  {journal} {Phys. Rev. B}\ }\textbf {\bibinfo {volume} {94}},\
  \bibinfo {pages} {241409} (\bibinfo {year} {2016})}\BibitemShut {NoStop}%
\bibitem [{\citenamefont {Slot}\ \emph {et~al.}(2017)\citenamefont {Slot},
  \citenamefont {Gardenier}, \citenamefont {Jacobse}, \citenamefont {van
  Miert}, \citenamefont {Kempkes}, \citenamefont {Zevenhuizen}, \citenamefont
  {Smith}, \citenamefont {Vanmaekelbergh},\ and\ \citenamefont
  {Swart}}]{slot2017}%
  \BibitemOpen
  \bibfield  {author} {\bibinfo {author} {\bibfnamefont {M.~R.}\ \bibnamefont
  {Slot}}, \bibinfo {author} {\bibfnamefont {T.~S.}\ \bibnamefont {Gardenier}},
  \bibinfo {author} {\bibfnamefont {P.~H.}\ \bibnamefont {Jacobse}}, \bibinfo
  {author} {\bibfnamefont {G.~C.~P.}\ \bibnamefont {van Miert}}, \bibinfo
  {author} {\bibfnamefont {S.~N.}\ \bibnamefont {Kempkes}}, \bibinfo {author}
  {\bibfnamefont {S.~J.~M.}\ \bibnamefont {Zevenhuizen}}, \bibinfo {author}
  {\bibfnamefont {C.~M.}\ \bibnamefont {Smith}}, \bibinfo {author}
  {\bibfnamefont {D.}~\bibnamefont {Vanmaekelbergh}}, \ and\ \bibinfo {author}
  {\bibfnamefont {I.}~\bibnamefont {Swart}},\ }\href
  {http://dx.doi.org/10.1038/nphys4105} {\bibfield  {journal} {\bibinfo
  {journal} {Nat. Phys.}\ }\textbf {\bibinfo {volume} {13}},\ \bibinfo {pages}
  {672} (\bibinfo {year} {2017})}\BibitemShut {NoStop}%
\bibitem [{\citenamefont {Drost}\ \emph {et~al.}(2017)\citenamefont {Drost},
  \citenamefont {Ojanen}, \citenamefont {Harju},\ and\ \citenamefont
  {Liljeroth}}]{drost2017}%
  \BibitemOpen
  \bibfield  {author} {\bibinfo {author} {\bibfnamefont {R.}~\bibnamefont
  {Drost}}, \bibinfo {author} {\bibfnamefont {T.}~\bibnamefont {Ojanen}},
  \bibinfo {author} {\bibfnamefont {A.}~\bibnamefont {Harju}}, \ and\ \bibinfo
  {author} {\bibfnamefont {P.}~\bibnamefont {Liljeroth}},\ }\href {\doibase
  10.1038/nphys4080} {\bibfield  {journal} {\bibinfo  {journal} {Nat. Phys.}\
  }\textbf {\bibinfo {volume} {13}},\ \bibinfo {pages} {668} (\bibinfo {year}
  {2017})}\BibitemShut {NoStop}%
\bibitem [{\citenamefont {Li}\ \emph {et~al.}(2016)\citenamefont {Li},
  \citenamefont {Qiu},\ and\ \citenamefont {Gao}}]{lishuai2016}%
  \BibitemOpen
  \bibfield  {author} {\bibinfo {author} {\bibfnamefont {S.}~\bibnamefont
  {Li}}, \bibinfo {author} {\bibfnamefont {W.-X.}\ \bibnamefont {Qiu}}, \ and\
  \bibinfo {author} {\bibfnamefont {J.-H.}\ \bibnamefont {Gao}},\ }\href
  {\doibase 10.1039/C6NR03223K} {\bibfield  {journal} {\bibinfo  {journal}
  {Nanoscale}\ }\textbf {\bibinfo {volume} {8}},\ \bibinfo {pages} {12747}
  (\bibinfo {year} {2016})}\BibitemShut {NoStop}%
\bibitem [{\citenamefont {Ropo}\ \emph {et~al.}(2014)\citenamefont {Ropo},
  \citenamefont {Paavilainen}, \citenamefont {Akola},\ and\ \citenamefont
  {R\"as\"anen}}]{dft2014}%
  \BibitemOpen
  \bibfield  {author} {\bibinfo {author} {\bibfnamefont {M.}~\bibnamefont
  {Ropo}}, \bibinfo {author} {\bibfnamefont {S.}~\bibnamefont {Paavilainen}},
  \bibinfo {author} {\bibfnamefont {J.}~\bibnamefont {Akola}}, \ and\ \bibinfo
  {author} {\bibfnamefont {E.}~\bibnamefont {R\"as\"anen}},\ }\href {\doibase
  10.1103/PhysRevB.90.241401} {\bibfield  {journal} {\bibinfo  {journal} {Phys.
  Rev. B}\ }\textbf {\bibinfo {volume} {90}},\ \bibinfo {pages} {241401}
  (\bibinfo {year} {2014})}\BibitemShut {NoStop}%
\bibitem [{\citenamefont {Kresse}\ and\ \citenamefont
  {Furthm\"uller}(1996)}]{vasp}%
  \BibitemOpen
  \bibfield  {author} {\bibinfo {author} {\bibfnamefont {G.}~\bibnamefont
  {Kresse}}\ and\ \bibinfo {author} {\bibfnamefont {J.}~\bibnamefont
  {Furthm\"uller}},\ }\href {\doibase 10.1103/PhysRevB.54.11169} {\bibfield
  {journal} {\bibinfo  {journal} {Phys. Rev. B}\ }\textbf {\bibinfo {volume}
  {54}},\ \bibinfo {pages} {11169} (\bibinfo {year} {1996})}\BibitemShut
  {NoStop}%
\bibitem [{\citenamefont {Perdew}\ \emph {et~al.}(1996)\citenamefont {Perdew},
  \citenamefont {Burke},\ and\ \citenamefont {Ernzerhof}}]{gga}%
  \BibitemOpen
  \bibfield  {author} {\bibinfo {author} {\bibfnamefont {J.~P.}\ \bibnamefont
  {Perdew}}, \bibinfo {author} {\bibfnamefont {K.}~\bibnamefont {Burke}}, \
  and\ \bibinfo {author} {\bibfnamefont {M.}~\bibnamefont {Ernzerhof}},\ }\href
  {\doibase 10.1103/PhysRevLett.77.3865} {\bibfield  {journal} {\bibinfo
  {journal} {Phys. Rev. Lett.}\ }\textbf {\bibinfo {volume} {77}},\ \bibinfo
  {pages} {3865} (\bibinfo {year} {1996})}\BibitemShut {NoStop}%
\bibitem [{\citenamefont {Tersoff}\ and\ \citenamefont
  {Hamann}(1983)}]{tersoff1983}%
  \BibitemOpen
  \bibfield  {author} {\bibinfo {author} {\bibfnamefont {J.}~\bibnamefont
  {Tersoff}}\ and\ \bibinfo {author} {\bibfnamefont {D.~R.}\ \bibnamefont
  {Hamann}},\ }\href {\doibase 10.1103/PhysRevLett.50.1998} {\bibfield
  {journal} {\bibinfo  {journal} {Phys. Rev. Lett.}\ }\textbf {\bibinfo
  {volume} {50}},\ \bibinfo {pages} {1998} (\bibinfo {year}
  {1983})}\BibitemShut {NoStop}%
\bibitem [{\citenamefont {Vicencio}\ \emph {et~al.}(2015)\citenamefont
  {Vicencio}, \citenamefont {Cantillano}, \citenamefont {Morales-Inostroza},
  \citenamefont {Real}, \citenamefont {Mej\'{\i}a-Cort\'es}, \citenamefont
  {Weimann}, \citenamefont {Szameit},\ and\ \citenamefont
  {Molina}}]{liebprl2015a}%
  \BibitemOpen
  \bibfield  {author} {\bibinfo {author} {\bibfnamefont {R.~A.}\ \bibnamefont
  {Vicencio}}, \bibinfo {author} {\bibfnamefont {C.}~\bibnamefont
  {Cantillano}}, \bibinfo {author} {\bibfnamefont {L.}~\bibnamefont
  {Morales-Inostroza}}, \bibinfo {author} {\bibfnamefont {B.}~\bibnamefont
  {Real}}, \bibinfo {author} {\bibfnamefont {C.}~\bibnamefont
  {Mej\'{\i}a-Cort\'es}}, \bibinfo {author} {\bibfnamefont {S.}~\bibnamefont
  {Weimann}}, \bibinfo {author} {\bibfnamefont {A.}~\bibnamefont {Szameit}}, \
  and\ \bibinfo {author} {\bibfnamefont {M.~I.}\ \bibnamefont {Molina}},\
  }\href {\doibase 10.1103/PhysRevLett.114.245503} {\bibfield  {journal}
  {\bibinfo  {journal} {Phys. Rev. Lett.}\ }\textbf {\bibinfo {volume} {114}},\
  \bibinfo {pages} {245503} (\bibinfo {year} {2015})}\BibitemShut {NoStop}%
\bibitem [{\citenamefont {Mukherjee}\ \emph {et~al.}(2015)\citenamefont
  {Mukherjee}, \citenamefont {Spracklen}, \citenamefont {Choudhury},
  \citenamefont {Goldman}, \citenamefont {\"Ohberg}, \citenamefont
  {Andersson},\ and\ \citenamefont {Thomson}}]{liebprl2015b}%
  \BibitemOpen
  \bibfield  {author} {\bibinfo {author} {\bibfnamefont {S.}~\bibnamefont
  {Mukherjee}}, \bibinfo {author} {\bibfnamefont {A.}~\bibnamefont
  {Spracklen}}, \bibinfo {author} {\bibfnamefont {D.}~\bibnamefont
  {Choudhury}}, \bibinfo {author} {\bibfnamefont {N.}~\bibnamefont {Goldman}},
  \bibinfo {author} {\bibfnamefont {P.}~\bibnamefont {\"Ohberg}}, \bibinfo
  {author} {\bibfnamefont {E.}~\bibnamefont {Andersson}}, \ and\ \bibinfo
  {author} {\bibfnamefont {R.~R.}\ \bibnamefont {Thomson}},\ }\href {\doibase
  10.1103/PhysRevLett.114.245504} {\bibfield  {journal} {\bibinfo  {journal}
  {Phys. Rev. Lett.}\ }\textbf {\bibinfo {volume} {114}},\ \bibinfo {pages}
  {245504} (\bibinfo {year} {2015})}\BibitemShut {NoStop}%
\bibitem [{\citenamefont {Park}\ and\ \citenamefont
  {Louie}(2009)}]{nanoletter2009}%
  \BibitemOpen
  \bibfield  {author} {\bibinfo {author} {\bibfnamefont {C.-H.}\ \bibnamefont
  {Park}}\ and\ \bibinfo {author} {\bibfnamefont {S.~G.}\ \bibnamefont
  {Louie}},\ }\href {\doibase 10.1021/nl803706c} {\bibfield  {journal}
  {\bibinfo  {journal} {Nano Lett.}\ }\textbf {\bibinfo {volume} {9}},\
  \bibinfo {pages} {1793} (\bibinfo {year} {2009})}\BibitemShut {NoStop}%
\bibitem [{\citenamefont {Polini}\ \emph {et~al.}(2013)\citenamefont {Polini},
  \citenamefont {Guinea}, \citenamefont {Lewenstein}, \citenamefont
  {Manoharan},\ and\ \citenamefont {Pellegrini}}]{polini2013}%
  \BibitemOpen
  \bibfield  {author} {\bibinfo {author} {\bibfnamefont {M.}~\bibnamefont
  {Polini}}, \bibinfo {author} {\bibfnamefont {F.}~\bibnamefont {Guinea}},
  \bibinfo {author} {\bibfnamefont {M.}~\bibnamefont {Lewenstein}}, \bibinfo
  {author} {\bibfnamefont {H.~C.}\ \bibnamefont {Manoharan}}, \ and\ \bibinfo
  {author} {\bibfnamefont {V.}~\bibnamefont {Pellegrini}},\ }\href
  {http://dx.doi.org/10.1038/nnano.2013.161} {\bibfield  {journal} {\bibinfo
  {journal} {Nat. Nanotechnol.}\ }\textbf {\bibinfo {volume} {8}},\ \bibinfo
  {pages} {625} (\bibinfo {year} {2013})}\BibitemShut {NoStop}%
\bibitem [{\citenamefont {Gibertini}\ \emph {et~al.}(2009)\citenamefont
  {Gibertini}, \citenamefont {Singha}, \citenamefont {Pellegrini},
  \citenamefont {Polini}, \citenamefont {Vignale}, \citenamefont {Pinczuk},
  \citenamefont {Pfeiffer},\ and\ \citenamefont {West}}]{west2009}%
  \BibitemOpen
  \bibfield  {author} {\bibinfo {author} {\bibfnamefont {M.}~\bibnamefont
  {Gibertini}}, \bibinfo {author} {\bibfnamefont {A.}~\bibnamefont {Singha}},
  \bibinfo {author} {\bibfnamefont {V.}~\bibnamefont {Pellegrini}}, \bibinfo
  {author} {\bibfnamefont {M.}~\bibnamefont {Polini}}, \bibinfo {author}
  {\bibfnamefont {G.}~\bibnamefont {Vignale}}, \bibinfo {author} {\bibfnamefont
  {A.}~\bibnamefont {Pinczuk}}, \bibinfo {author} {\bibfnamefont {L.~N.}\
  \bibnamefont {Pfeiffer}}, \ and\ \bibinfo {author} {\bibfnamefont {K.~W.}\
  \bibnamefont {West}},\ }\href {\doibase 10.1103/PhysRevB.79.241406}
  {\bibfield  {journal} {\bibinfo  {journal} {Phys. Rev. B}\ }\textbf {\bibinfo
  {volume} {79}},\ \bibinfo {pages} {241406} (\bibinfo {year}
  {2009})}\BibitemShut {NoStop}%
\end{thebibliography}%

\end{document}